# EFFECTS OF EARLY INTENSE BOMBARDMENT ON MEGAREGOLITH EVOLUTION AND ON LUNAR (AND PLANETARY) SURFACE SAMPLES

William K. Hartmann[1] and Alessandro Morbidelli[2]

[1]Planetary Science Institute, Tucson, Arizona

[2]Université Côte d'Azur, CNRS–Lagrange

Observatoiré de la Côte d'Azur, CS 34229–F 06304 Nice Cedex 4, France

**ABSTRACT** — Impact rates in the first 500 Myr of the solar system are critical to an understanding of lunar geological history, but they have been controversial. The widely accepted, post-Apollo paradigm of early lunar impact cratering (ca. 1975-2014) proposed very low or negligible impact cratering in the period from accretion (>4.4 Ga) to ~4.0 Ga ago, followed by a ~170-million-year-long spike of cataclysmic cratering, during which most prominent multi-ring impact basins formed at age ~3.9 Ga.  More recent dynamical models suggest very early intense impact rates, declining throughout the period from accretion until an age of ~3.0 Ga.  These models remove the basin-forming spike.  This shift has important consequences *vis-à-vis* megaregolith evolution and properties of rock samples that can be collected on the lunar surface today.  We adopt the Morbidelli et al. (2018) "accretion tail" model of early intense bombardment, declining as a function of time.  We find effects differing from the previous models: early crater saturation and supersaturation; disturbance of magma ocean solidification; deep early megaregolith; and erosive destruction of the earliest multi-ring basins, their impact melts, and their ejecta blankets.  Our results explain observations such as differences in numbers of early lunar impact melts vs. numbers of  early igneous crustal rocks,  highland breccias containing impact melts as old as 4.35 Ga, absence of a 170 Myr-long spike in impact melt ages at 3.9 Ga among lunar and asteroidal meteorites, and GRAIL observations of lunar crustal structure.

## BACKGROUND

Early analysis of Apollo and Luna samples from the Moon, ca. 1970s, showed an unexpected paucity of rocks older than about 4.0 Ga, especially among impact melts.  This quickly led to suggestions that many or most of the front-side multi-ring basins formed in a narrow time interval around 3.9 Ga ago (e.g., Tera et al. 1973; Turner and Cadogan 1973).  Other interpretations were suggested, but this "terminal cataclysm" was strengthened when Ryder (1990), focusing on Apollo impact melt rock samples, found a narrow, ~170 Ma-wide spike in impact melt dates around 3.85 to 4.00 Ga, with few impact melts before that.  Ryder argued that lack of impact melts meant lack of impacts, and his work solidified what can fairly be called a four-decade paradigm, according to which the post-accretion Moon experienced negligible impacts from about 4.4 to 4.0 Ga, followed by a "terminal cataclysm" (also called "Late Heavy Bombardment" or "LHB"), during which the major multi-ring basins of the front side were formed.  For brevity, we will call this interpretation the "classic" version of the terminal cataclysm/LHB paradigm.  In short, the early post Apollo period, marked widespread acceptance of the "classic" terminal cataclysm/LHB scenario.

Contrary data began to appear but were not always recognized as such. For example, KREEP-poor impact melt clasts in lunar meteorites showed no spike at 3.9 Ga, and also showed a paucity of impact melts before ~4.0 Ga, but these valuable data were said to support the terminal cataclysm model in spite of the missing spike (Cohen et al. 2000). Dynamical models, such as the original "Nice model" (Gomes et al. 2005) and models by Bottke et al. (2010, 2012) were presented as support or confirmation for the "classic" terminal cataclysm paradigm, because they provided a conceptual explanation for the possible existence of a sudden surge of impactors at 3.9 Ga, associated with a phase of dynamical instability of the giant planets. The timing of the spike at 3.9 Ga in those models was not a direct consequence of the dynamical evolution and (as stated in the models) was assumed. These models implied that the sharp spike in cratering applied throughout the solar system at 3.9 Ga, but this was not confirmed in asteroidal meteorites (e.g., see data related to Vesta, Cohen 2013).

Dynamicists continued to use the term "LHB" but changed the definition from a sharp spike lasting ~170 million years (Ryder 1990) to a gentle surge lasting as much as 1600 million years (Bottke et al. 2012). This maintained the "LHB" terminology in spite of modifying the phenomenon. More recently, it has been realized that lack of early impact melts may not require lack of impacts, but rather may simply mean lack of *survival* of impact melts — due either to lack of impacts or destruction of the early specimens (Hartmann 1975, 2003, 2019; Chapman et al. 2007). Morbidelli et al. (2018) compared an updated "classic" paradigm dynamical model (which they refer to as a "cataclysm scenario") to a second, early intense bombardment model with a monotonically decreasing impact rate, which they call the "accretion tail scenario." The accretion tail model is similar to some of the early early intense bombardment models with decreasing impact rate, based on planetary accretion theory and crater counts (Safronov 1972; Hartmann 1966, 1970a,b, 1972, 1975; Turner 1979; Neukum 1983. Those models, however, lacked detailed dynamical modeling.

One problem with the early versions of the accretion tail scenario is that, in order to cause the proposed impact rates around 4 Ga, the total amount of mass accreted by the Moon since its formation would have exceeded by one order of magnitude that inferred by some researchers from the concentration of highly siderophile elements (HSEs) in the lunar mantle (Day et al. 2007; Day and Walker 2015). However, Morbidelli et al. (2018) suggested that HSEs should have been sequestered into the lunar core during mantle crystallization and overturn. If the latter happened around 4.35 Gy ago, as predicted by lunar thermal evolution models (Elkins-Tanton et al. 2011) and supported by the ages of the oldest lunar rocks (Borg et al. 2015), an intense early bombardment followed by the accretion tail scenario could also have been an influence on HSE concentration in the current lunar mantle and is consistent with the cratering rate observed on terrains younger than 4 Ga. Zhu et al. (2019) supported this result by computing the amount of accreted material implanted in the lunar crust and mantle in each impact of the accretion tail scenario. (See also Joy et al. 2016 for discussion of the types of meteorites striking the Moon at different times.) Based on such work, Morbidelli et al. (2018) concluded that the accretion tail scenario is preferable to the terminal cataclysm scenario, because it avoids the need to delay a giant planet orbit instability until 3.9 Ga — a delay that has been suggested to be inconsistent with the probable dynamics of giant planet formation (Ribeiro de Sousa et al. 2020). Similarly,

Nesvorny et al. (2019) found evidence that the trans-Neptunian disk of planetesimals (scattered in order to produce an impact spike in the inner solar system) should have been dynamically depleted within 100 Ma of solar system origin, consistent with the growing evidence that a sharp spike at 3.9 Ga did not occur. According to such studies, the maximum cratering rates in the inner solar system should have occurred much earlier than 3.9 Ga (Clement et al. 2018), possibly helping to explain the giant Moon-forming impact at a very early date >4.4 Ga. In short, the present paper will examine the lunar geological consequence of the early intense bombardment/accretion tail scenario that we infer from the apparent collapse of the "classic" cataclysm/LHB paradigm.

The need for our present work is shown by the fact that, as mentioned above, numbers of authors have continued to use the term "LHB" in spite of the changing definitions (sometimes from year to year), while other authors hold to the original definition of a spike at 3.9 Ga. (See Hartmann 2019 for more detailed discussion.) To emphasize the importance of the definitions, we note that in spite of the shift away from the terminal cataclysm/LHB model, the scientific and popular literature in various fields has continued to appeal to the "classic" version of the terminal cataclysm or LHB as well established. For example:

- Perkins (2014, writing in *Science*, reviewing how cosmic impacts could have transformed simple precursor materials into nucleobases in RNA): "During a period aptly dubbed the Late Heavy Bombardment, which began about 4 billion years ago and lasted some 150 million years…."
- Jolliff and Robinson (2019, in *Physics Today*, reviewing the legacy of Apollo): "The Apollo samples provided the first evidence of the so-called late, heavy bombardment…thought to have spiked around 3.9-4.0 billion years ago. Models of the early solar system's orbital dynamics suggest that shifts in the orbits of Jupiter and Saturn may have destabilized early asteroid and cometary belts and led to that cataclysm some 500 million years after the solar system formed."
- Cartier (2019, in *Eos*, writing about two zircon grains found in a lunar meteorite): "Past studies have shown that these two lunar zircons are about 4.1 billion years old and formed during a tumultuous time in the solar system's history called the Late Heavy Bombardment."
- Shigeru Ida (2019, in *Science*, writing about Saturn's rings' origin possibly involving breakup of an incoming Kuiper Belt object): "Such an encounter would have been very rare except during the Solar System formation stage…and the Late Heavy Bombardment era at ~4 billion years ago."

These continued applications of the cataclysm/LHB concept in fields as diverse as biology, lunar history, meteorite evolution, and Saturn rings origin, show the need to reassess the inner solar system cratering from ages of ~ 4.4 Ga to ~ 3.8 Ga.

A note about our terminology: The present paper will deal with various kinds of intervals of years, not only conventional "age," (measured back in time from today), but also intervals measured forward in time from the formation of the Moon, and simple intervals such as the duration of magma ocean solidification, or the ~170 million year interval centered around 3.9 billion years ago that was the reported duration of the terminal cataclysm or LHB, based on impact melt ages (Ryder 1990). In the interest of clarity and brevity, we adopt a current convention (e.g., Christie-Blick 2012) in which conventional age is designated with "a" (Latin *annum* for year, hence Ga and Ma), but intervals, which are not ages, are designated with "yr." (Hence in this paper the reader fill find use both Ma and Myr).

Another clarification of the present paper may be helpful. Various readers of earlier drafts suggested that we add discussions of certain current issues along with references to various papers regarding the history of the terminal cataclysm/LHB paradigm. These issues included discussions of whether "paradigm" is an appropriate term, and more discussion of certain papers. Examples of such issues are more discussion of the useful work of Haskin (1998), Petro and Pieters (2008), and others about Imbrium as a source of 3.9 Ga impact melts and widely distributed ejecta; the role of Orientale basin as "abruptly" ending big-basin formation; alterations of the original Nice model with redefinitions of "LHB;" more discussion of various papers in the first decades of lunar sample analysis; etc. Because all of these issues were treated at length by one of us in a recent, open-access study of the origins, acceptance, and eventual rejection of the terminal cataclysm/LHB scenarios (78 pages with 181 references, Hartmann 2019), we beg the readers' indulgence in our referencing of that paper, rather than lengthening this paper by repeating those discussions here. We focus here on our main question: *If* the quantitative Morbidelli et al. (2018) "accretion tail" model of lunar history is correct, how does that affect the evolution of megaregolith and how does megaregolith evolution affect rock samples on the lunar surface today and, by extension, other planetary surfaces?

**ANALYSIS AND APPLICATION OF ACCRETION TAIL SCENARIO**
***VIS-À-VIS* IMPACT RATE VS. TIME**

As pointed out by Morbidelli et al. (2018), the "accretion tail" dynamical model creates a curve of declining impact rate vs. time, similar to — but independent of — early pre- and post-Apollo suggestions about the nature of that curve, based mainly on cratering statistics. Hartmann (1965, 1966) estimated the characteristic age of lunar *mare* lava plains at ~3.6 Ga, using cratering rates derived from the Canadian shield. The second paper pointed out that since the pre-mare lunar highlands had 32 × the crater density of the maria, the brief pre-3.6 Ga cratering rate had to average at least ~160× the post-mare crater rate. The uncertainty in that number was estimated to be a factor 2. The number 160 was considered a lower limit, because the factor 32 crater density was found on a number of solar system surfaces, indicating that it is represents the crater saturation equilibrium level found on many bodies (Hartmann 1984). Craters formed in excess of the saturation level are destroyed by later cratering, and this means that the total impact crater numbers in the highlands may be much larger than what we can see today. The saturation equilibrium level is important in the present paper because, as we will show, the total number of impacts during very early intense bombardment, based on Morbidelli et al. (2018), was well beyond the number required to reach saturation. Thus, the saturation curve found in most areas today does not represent the actual number of craters that were formed, but only a lower limit. Once saturation was exceeded on a surface, we lose information on the impact numbers and the shape of the earlier size-frequency distribution curve.

After investigators dated Apollo mission samples collected from various landing sites, crater density measurements as a function of measured surface age indicated that as we go backward in time from ages ~3.2 to ~3.8 Ga, the cratering rate reached a value of order ≥200 times the current rate (e.g., Hartmann 1972; Neukum 1983; Neukum et al. 2001). In the "accretion tail" scenario, this decline in cratering can be seen as the tail end of the more or less monotonic decline in impact rate from >4.4 Ga ago to 3.8 Ga ago, when the cratering record does not reach saturation, and is more clearly measurable. We stress that the time evolution of the impact rate in this paper not a backward extrapolation of the observed impact rate evolution observed in the 3.2-3.8 Ga interval, but is the result of the numerical modelling of the dynamical evolution of the planetesimals remaining in the terrestrial planet region at the end of the planets' formation process (Morbidelli et al. 2018).

Table 1 shows an application of this model. Column 1 lists dates at 100 Myr intervals. Columns 2 and 3 illustrate two ways of normalizing the bombardment model data to give cratering rates relative to the modern cratering rate (defined as 1.0). In column 2 (model A) we normalize to an average value of 1.0 for the last 400 Ma (averaging over the Morbidelli et al. 2018 model values for the last four 100-Ma bins, including T=0=today). Column 3 shows a second, different normalization of the curve (model B). It applies suggestions that the cratering rate at 3.0 Ga was probably several times that observed today, and was still declining from 3.0 Ga till today (Quantin et al. 2004, 2007; Hartmann et al. 2007). Thus, column 3 normalizes the data to a value of 3.0 at 3.0 Ga. The two normalizations give similar results, as plotted in Fig. 1.

In this context, it is important to note that the Morbidelli et al. (2018) data set, in order to extend the timeline from 3.0 Ga until today, involves a conceptual discontinuity at ~3.0 Ga ago. At age >3 Ga it follows the calculated decay of asteroids and leftover planetesimals assuming their dynamics were dominated by planetary perturbations (the main subject of Morbidelli et al. 2018). For age <3 Ga, the model shifts to dominance of near-Earth asteroids escaping from the asteroid belt under Yarkovsky forces, giving a relatively constant bombardment in Table 1 (similar to the normalization in column 2 of Table 1). In reality (especially with an asteroid belt a bit more massive 3 Ga ago than now, giving more NEAs) the two curves would merge more smoothly. For this reason, Morbidelli et al. (2018, their Fig. 5) originally plotted their results from 4.5 Ga only to 3.0 Ga ago, but here we use additional data from their model to extend the curve.

The fourth column in Table 1 shows a few examples of earlier cratering rate estimates (still normalized to current rate = 1) based on extrapolation of crater count data backward to ~4.3 or 4.4 Ga, assuming the early estimates of decline in planetesimal numbers (Hartmann 1972; Neukum et al. 2001; Quantin et al. 2007). In Table 1, fluctuations in cratering rate and half-lives of asteroids against collisions with other bodies, at ages around 3.0 Ga may be due mostly to the merger of the two models. The final (fifth) column of Table 1 lists some estimates of half-lives of the projectile population published during earlier crater-count studies (Hartmann 1972; Neukum et al. 2001; Quantin et al. 2007). We note that these are similar to the results from Morbidelli et al. (2018).

Figure 1 shows the Morbidelli at al. (2018) cratering rate curve as a function of time (on a logarithmic scale), using the two normalization methods from Table 1, allowing either no decline since 3.0 Ga or decline by a factor 3. We infer that the most likely cratering rate behavior occurs between these two curves, but with a more gradual transition at about 3.5 to 2.5 Ga, ending (by definition) with today's rate of 1.0.

To summarize, the "accretion tail" scenario favored by Morbidelli et al. (2018) indicates that the early intense bombardment rates in the first few hundred million years (surface formation until ~4.15 Ga) exceeded 1000 times the present rate. At dates before 4.4 Ga, it may have exceeded 10,000 the current rate, although, as we probe back toward the beginning, the lunar surface may have been in magma-ocean state. Note the contrast with the "classic" terminal cataclysm paradigm, which called for more or less negligible cratering in the ~4.4 Ga to ~4.0 age interval, with few if any of the major front-side multi-ring basins forming in that period.

**Table 1.** Cratering rate relative to present impact rate based on Morbidelli et al. (2018) accretion tail scenario

| Age (time before present, Ga) | Cratering rate relative to present (Model A = constant since 3.0 Ga) (Morbidelli et al. ×1.06) | Cratering rate relative to present (Model B = decline by factor 3 since 3.0 Ga) (Morbidelli et al. ×1.58) | Examples of cratering rate relative to present estimated from crater-chronometry techniques* | Half-life estimates of the projectile population* |
|---|---|---|---|---|
| 4.5 | 33,600 | 50,100 | | ~20 Myr (M) |
| 4.4 | 12,900 | 19,200 | | ~50 Myr (M) |
| 4.3 | 3500 | 5200 | | ~60 Myr (M) |
| | | | | ~80 Myr (H) |
| 4.2 | 1272 | 1900 | | ~70 Myr (M) |
| 4.1 | 550 | 815 | Average of ≥160 before 4.0 Ga (Hartmann 1966) | ~90 Myr (M) |
| 4.0 | 260 | 390 | 500 (Neukum et al. 2001, p. 69) | ~95 Myr (M) |
| 3.9 | 140 | 209 | | ~100 Myr (M) |
| 3.8 | 66 | 98 | | ~120 Myr (M) |

| Age (time before present, Ga) | Cratering rate relative to present (Model A = constant since 3.0 Ga) (Morbidelli et al. ×1.06) | Cratering rate relative to present (Model B = decline by factor 3 since 3.0 Ga) (Morbidelli et al. ×1.58) | Examples of cratering rate relative to present estimated from crater-chronometry techniques* | Half-life estimates of the projectile population* |
|---|---|---|---|---|
| 3.7 | 40 | 60 | | ~130 Myr (M) |
| 3.6 | 30 | 44 | | ~140 Myr (M) |
| | | | | ~ 300 Myr (H) |
| 3.5 | 15 | 22 | | ~140 Myr (M) |
| 3.4 | 9.4 | 14 | | ~160 Myr (M) |
| 3.3 | 6.3 | 9.3 | | ~180 Myr (M) |
| 3.2 | 4.2 | 6.3 | | ~230 Myr (M) |
| 3.1 | 3.9 | 5.8 | | ~210 Myr (M) |
| 3.0 | 2.0 | 3.9 | ~2 to ~4× higher than present rate (Quantin et al. (2007) | ~120 Myr (M) |
| 2.9 | 0.95 | 1.4 | | ~175 Myr (M) |
| 2.8 | 0 95 | 1.4 | ~1.0 from 2.8 Ga till today (Neukum 1983) | >500 Myr (N) |
| 2.7 | 1.1 | 1.6 | | >500 Myr (M,N) |
| 2.6 | 1.3 | 1.9 | | >500 Myr (M,N) |
| 2.5 | 0.95 | 1.4 | | >500 Myr (M,N) |
| | | | | ~1000 Myr (Q) |

*Half-life estimates: M = Morbidelli et al. (2018) dynamical model; N = Neukum 1983; H = Hartmann 1972, Fig. 3, p. 55; Q = Quantin et al. 2007). Note: Morbidelli et al. (2018) are given precedence in this paper over the earlier estimates, which are cited merely to demonstrate similarity with early post-Apollo models based on crater statistics.

As Morbidelli et al. (2018) reported, the crater count results (and their extrapolation to before 4.0 Ga) come strikingly close to the results from the Apollo-era models in both shape and absolute numbers. For example, Hartmann (1972), using early Apollo sample data stated that prior to 4.1 [Ga] ago, the cratering rate on the Moon was at least $10^3$ times the present rate and the rate declined with a half-life less than $8 \times 10^7$ yr. During the interval from 4.1 to 3.2 [Ga] ago, the number of planetesimals showed an exponential decay with a half-life about $3 \times 10^8$ yr…. A more constant cratering rate applied in the last three aeons.

Neukum (1983, as reviewed in Neukum et al. 2001, their Fig. 11) developed an analytic interpretation of the crater-count data, an equation fitting their observations of crater cumulative crater density as a function of time. The differential values of that equation (i.e., difference in craters/km$^2$ in a specified size range from beginning to end of 100 Myr intervals), give the crater formation rate (i.e., craters/km$^2$ in a specified size range) formed per 100 Myr interval. As an example of their results, Neukum et al. (2001) stated that "…the cratering rate 4 Gyr ago was 500 times higher than the constant rate during the last 3 Gyr," a statement close to the curves in our Fig. 1.

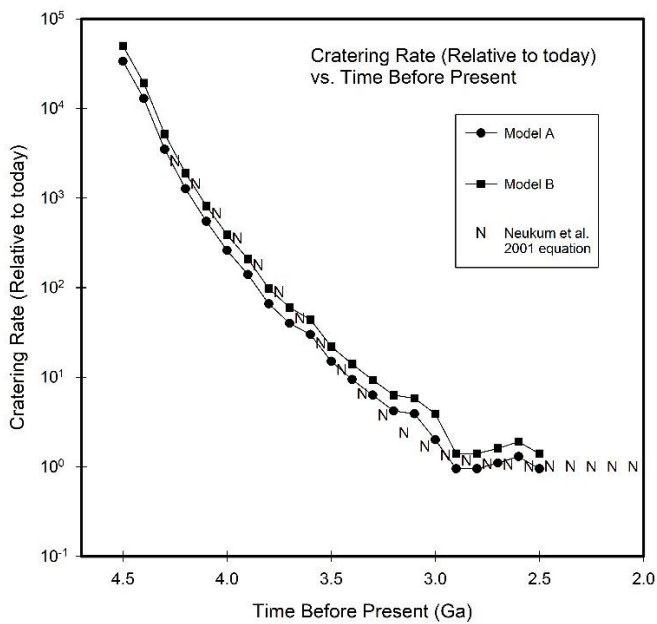

**Figure 1.** Curve showing cratering rate as a function of time (relative to today's cratering rate) derived from Morbidelli et al. (2018) accretion tail model. Lower curve assumes cratering rate asymptotically arrives at today's rate (i.e., 10°). Upper curve assumes cratering rate at 3.0 Ga equals 3 × today's rate. "N" symbols are independent data from crater counts, taken from equation fitted by Neukum to crater count data (see text for further discussion).

The impact rate data points derived from the Neukum equation, based entirely on crater counts (not on a dynamical model) are included with symbol "N" for comparison in our Fig. 1. The agreement between the dynamical models A and B, and the cratering rate analysis, is remarkable for three distinct reasons. (1) The research teams are entirely independent. The Morbidelli et al. (2018) team made no attempt to match their dynamical model to the earlier crater chronometry results. (2) The techniques used are entirely independent; mathematical results from orbital dynamical calculations are unrelated in practice to counts of lunar craters. (3) Perhaps most striking, Morbidelli et al. (2018) model predictions match not only the best-established parts of the Neukum curve at ages ≲3.8 Ga, but also an extrapolation of that curve to earlier dates. To repeat, the best data used for the Neukum curve come from dated samples returned from Apollo and Luna landing sites, but surfaces for crater counts dating from before ~3.8 Ga (a) are less well documented (if available at all), and (b) have crater densities near the saturation equilibrium line so that super-saturation densities on older surface cannot be directly measured. All the more noteworthy, then, that the Neukum equation for cratering rates fits so well the results of the dynamical model.

The conclusion of this section is that the accretion tail model, along with certain Apollo era suggestions of early intense cratering, imply that the cratering rate prior to 3.8 or 3.9 Ga likely exceeded hundreds and even thousands of times the rate from the last 1-3 Gyr (and could plausibly have exceeded ten thousand times higher than it is today, prior to ~4.35 Ga), as shown in Table 1, columns 2 and 3, from the Morbidelli et al. (2018) data. This is contrary to "classic" terminal cataclysm/LHB paradigm, where cratering rates prior to 3.9-4.0 Ga were considered to be comparable to the present rate. It is significant that, half a century after the Apollo and Luna samples were returned, the accuracy of first significant figures is an issue in understanding important lunar phenomena.

## DEVELOPING TOOLS TO STUDY EFFECTS
## OF INTENSE EARLY BOMBARDMENT

A useful tool in understanding the lunar cratering record is the "isochron diagram," which shows the observed size-frequency distribution (SFD) of impact craters as a function of surface age. In the template version, shown here in Fig. 2a, the crater density (craters/km$^2$) is shown incrementally, for each individual size bin, and the size bins are plotted on a log$_{\sqrt{2}}$ scale, for example from 1.0 km to 1.4 km, from 1.4 km to 2 km, and so on. This plot has an advantage over widely used cumulative plots, because it shows exactly what is happening in each individual size bin, rather than smoothing the curve's structure over many sizes, as cumulative curves do.

The isochron diagram has a number of additional useful features. The upper diagonal boundary, marked "Saturation" shows an empirical curve (mentioned earlier), which has been repeatedly observed in the most heavily cratered surfaces on the Moon, various asteroids, and outer planet satellites throughout the solar system (Hartmann 1984). The observed saturation curve is not a "geometric" saturation, since more small craters could, geometrically, be squeezed into existing empty spaces on these surfaces. However, nature does not allow the addition of craters in only a narrow size range; the whole SFD must be added all at once, which requires occasional, large, basin-scale impacts, whose excavation and ejecta blankets completely resurface some areas. Furthermore, "equilibrium" connotes stability, whereas empirical saturation in a given region does not represent a static situation. Not only do the distant (often "off-stage") basins' ejecta occasionally wipe out shallower craters (at, say, D <100 m diameter, depending on distance and size of the basin), but also the continual rain of meteorites creates craters smaller than any specified diameter D, and erodes rims of craters larger than size D by a simple sandblasting effect. An inexorably increasing supersaturation would produce many big, deep impact craters (some no longer visible) and thus pulverize increasingly deep layers, including the earliest impact melt lenses formed in the near-surface kilometers under the earliest large impact basin floors. Thus, different parts of the saturation line can thus oscillate over time by factors about 2 to 4 (Hartmann and Gaskell 1997).  The dashed lines on either side of the saturation curve thus show the typical range of oscillation of visible saturation crater densities.  The important larger lesson is that a surface characterized by the saturation line may have been impacted far more times than is evident from the numbers of visible craters, preventing a clear measurement of the SFD of the earliest impactors.

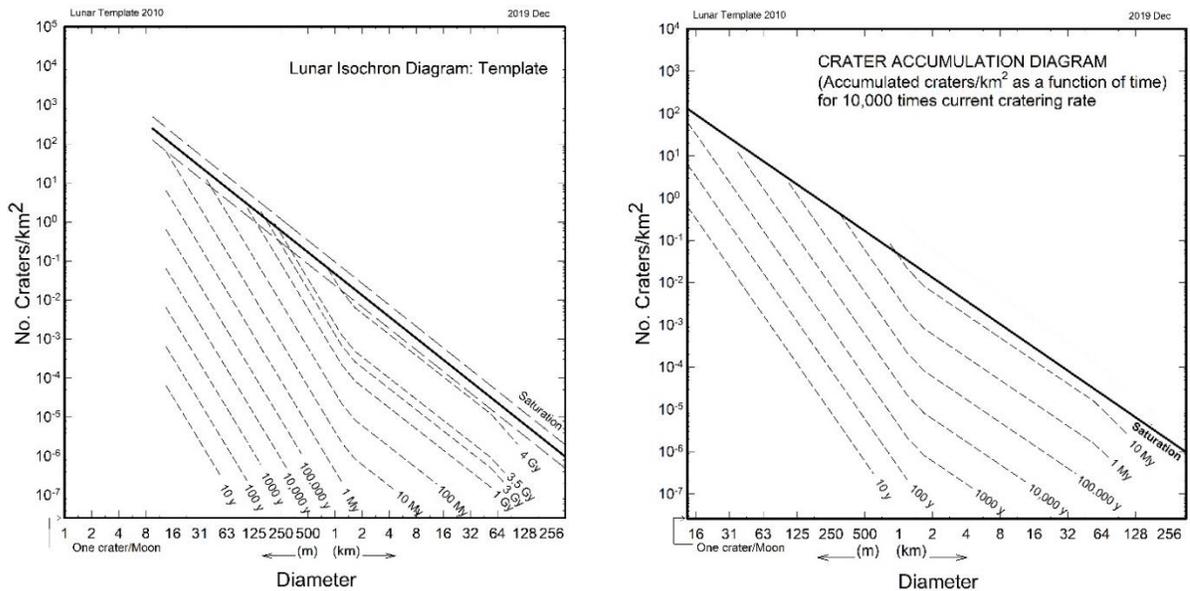

**Figure 2.** (a)  Template for lunar isochron diagram.  Each curve shows the size-frequency distribution of craters observed on lunar surfaces of specified ages.  The diagram assumes constant crater production rate going back in time until about 3 Ga but higher rates before that, as based on Apollo data (Hartmann 1970a,b, 1972; Neukum et al. 2001). (b) Crater accumulation diagram under conditions of impact rate 10,000 times the current rate, as suggested by Morbidelli et al. (2018) for the period around 4.35 Ga.  Given the half-life estimates in Table 2, a rate of ~ 10,000× current rate lasts only tens of millions of years, so the graph is extended only to 10 Myr (see text for further discussion).  Because crater chronometry is constantly improving with new data, fine print above diagrams refers to date of our latest iteration of the lunar isochron plot, and the date of our data set.

     Now suppose the cratering rate ~4.35 Ga was ~10,000 times the present rate, as the accretion tail model indicates for an interval of about 50-100 Myr, centered on that date (see Table 1 and Fig. 1).  To create an isochron diagram to match conditions under that elevated impact rate, we simply relabel the isochrons, stopping at 10 Myr, because the half-life of decline in cratering rate at 4.35 would have been on the order of 50-60 Myr, according to Table 1.  The result is shown in Fig. 2b.  The diagram suggests that any surface formed during cratering at that rate would have reached supersaturation equilibrium conditions on a timescale ~50 Ma, and we can surmise that multi-ring basins of D >300 km would have been forming in stochastic intervals averaging 50-100 Myr.  If the cratering rates were higher than 10,000 times the present rate

before 4.35 Ga (as indicated in Table 1 and Fig. 1), and if the surface were solid, supersaturation would have been reached in <50 Ma.

## CRATER ACCUMULATION ON SURFACES FORMED AT DIFFERENT TIMES

As suggested by Fig. 2b and the discussion in the last section, the accretion tail model offers an approach to understanding how the earliest cratering conditions affected the early lunar crust and megaregolith, and even the kinds of samples picked up on the surface today. This investigation is detailed in Table 2.

**Table 2.** Calculation of early crater accumulation. (Deriving the accumulated number of craters/km² starting at specified date, expressed in terms of average mare density* = 1.)

| Age (Ga) | ΔN(1) 1 km craters formed per km² during 100 Myr interval specified in column 1 | ΔN(1) relative to average mare, where N(1) crater density [= 3.2 (10⁻³)]* | Start at 4.55 Ga (if Moon existed) | Start at 4.45 Ga | Start at 4.35 Ga | Start at 4.25 Ga | Start at 4.05 Ga | Start at 3.85 Ga | Start at 3.65 Ga | Start at 3.45 Ga |
|---|---|---|---|---|---|---|---|---|---|---|
| 4.55 | | | 0 | | | | | | | |
| 4.5 | 31.7 | 9906 | | | | | | | | |
| 4.45 | | | 9906 | 0 | | | | | | |
| 4.4 | 1.21 | 378 | | | | | | | | |
| 4.35 | | | 10280 | 378 | 0 | | | | | |
| 4.3 | 0.33 | 103 | | | | | | | | |
| 4.25 | | | 10380 | 481 | 103 | 0 | | | | |
| 4.2 | 0.12 | 37 | | | | | | | | |
| 4.15 | | | 10420 | 518 | 140 | 37 | | | | |
| 4.1 | 0.052 | 16.1 | | | | | | | | |
| 4.05 | | | 10440 | 534 | 156 | 53 | 0 | | | |
| 4.0 | 0.024 | 7.62 | | | | | | | | |
| 3.95 | | | 10440 | 541 | 164 | 61 | 7.62 | | | |
| 3.9 | 0.013 | 4.13 | | | | | | | | |
| 3.85 | | | | 545 | 168 | 65 | 11.8 | 0 | | |
| 3.8 | 6.2 (-3) | 1.94 | | | | | | | | |
| 3.75 | | | | | 170 | 67 | 13.8 | 1.94 | | |
| 3.7 | 3.8 (-3) | 1.19 | | | | | | | | |
| 3.65 | * | | | | | 68 | 14.8 | 3.13 | 0 | |
| 3.6 | 2.3 (-3) | 0.718 | | | | | | | | |

| Age (Ga) | ΔN(1) 1 km craters formed per km² during 100 Myr interval specified in column 1 | ΔN(1) relative to average mare, where N(1) crater density [= 3.2 (10⁻³)]* | Start at 4.55 Ga (if Moon existed) | Start at 4.45 Ga | Start at 4.35 Ga | Start at 4.25 Ga | Start at 4.05 Ga | Start at 3.85 Ga | Start at 3.65 Ga | Start at 3.45 Ga |
|---|---|---|---|---|---|---|---|---|---|---|
| 3.55 | | | | | | | | 15.4 | 3.85 | 0.72 |
| 3.5 | 1.4 (-3) | 0.475 | | | | | | | | |
| 3.45 | | | | | | | | 16.0 | 4.33 | 1.20 | 0 |
| 3.4 | 8.9 (-4) | 0.278 | | | | | | | | |
| 3.35 | | | | | | | | | 4.61 | 1.48 | 0.28 |
| 3.3 | 5.9 (-4) | 0.184 | | | | | | | | |
| 3.25 | | | | | | | | | 4.79 ** | 1.66 ** | 0.46 ** |
| 3.2 | 4.0(-4) | 0.126 ** | | | | | | | | |
| 3.15 | | | | | | | | | 4.92 ** | 1.79 ** | 0.59 ** |
| 3.1 | 3.7(-4) | 0.116 ** | | | | | | | | |
| 3.05 | | | | | | | | | | 1.91 ** | 0.71 ** |
| 3.0 | 2.3(-4) | 0.072 ** | | | | | | | | |
| 2.95 | | | | | | | | | | | 0.78 ** |

*Average mare N(1) density is assumed to be 3.2(10⁻³)craters/km², larger than 1 km, based on Hartmann (1966, 2005) crater counts on a variety of lunar mare surfaces, which were found to have virtually the same crater density, independent of the Morbidelli et al. (2018) model. See text.

**The accumulated crater numbers in the last bins of the table should probably be higher. This involves intersection of two different processes of scattering interplanetary bodies in the Morbidelli et al. (2018) model, around 3 Ga. (See text for discussion.)

In Table 2, column 1 defines 100 Myr-long time bins, and lists dates of surface formation. In column 2, N(1) refers to the cumulative number of craters/km² that have diameter >1 km. N(1) is often used as a proxy for the entire size-frequency distribution curve for all impact craters, from microscopic pits to 1000-km multi-ring basin structures.  We adopt it as a first-order proxy, but we recommend caution, because the N(1) values fall just in the diameter zone where the -2 slope SFD of larger craters intersects the steeper -3 to -4 slope at smaller diameters, as shown in Fig. 2.  Our work concentrates on the rate of change in the entire crater population as a function of time, but N(1) still has first-order value in our discussions.  Thus, in column 2, ΔN(1) values from the accretion tail model are defined as the total number of craters with diameter D >1 km added per $km^2$ in the 100 Ma intervals shown in column 1. For example, from 3.05 to 2.95 Ga, centered on 3.0 Ga, the model shows ΔN(1) addition of ~2.3 ($10^{-4}$) craters/$km^2$ larger than 1 km.

Column 3 repeats column 2 by listing the ΔN(1) number of craters in terms of average mare crater density.  This gives a more direct connection to the appearance of the lunar surface, making the table more accessible to readers not familiar with N(1).  Entries of ~1.0 in column 2 thus represent a surface with crater densities typical of average lunar mare on the front side of the Moon, based partly on data from the University of Arizona Lunar and Planetary Laboratory crater catalogs (for a summary of those catalogs, see Wood 1972) and additional counts over the years, by Hartmann, treating the mare surface as one giant mare.  The mare surfaces in the major basins have remarkably similar densities, and "average lunar mare," as mentioned in a footnote to Table 2, thus refers to a summation of counts in each diameter bin from surfaces of various major basins, divided by the total area counted in each bin.

The remaining columns, 4 to 11, give crater densities relative to average mare crater density, as they accumulate on surfaces of eight difference starting dates.  Thus for example, if we look at a surface (perhaps a lava flow) formed at 4.35 Ga (column 6), it has 0 crater density at 4.35, but adds ΔN(1) of 103 craters/$km^2$ by 4.25 Ga.  And so on, down the table until we see it leveling off at roughly 170 craters/$km^2$ larger than 1 km, which would be observed today.  In columns 4 to 11, entries of ~32 represent lunar highland surfaces, matching saturation equilibrium as measured by Hartmann (1984).

Figure 3 shows a plot of the data in Table 2.  Three vertical scales have been added to the diagram to demonstrate its range of important information.  At the left side we see the crater densities in terms of average lunar mare and also in terms of N(1) as used in Neukum (1983) cumulative crater densities.  At the right side we see a listing of the estimated megaregolith depths, which are a direct function of accumulated crater densities, as will be discussed below. The shallower depth estimates come from Apollo observations and GRAIL data (Wieczorek et al. 2013), and the deeper depths, in the regions dating to the first 500 Myr, come from our calculations.  Figure 3 also contains three horizontal shaded bands, marking average lunar mare crater densities, saturation density (32× average mare density), and 10× saturation density.  The vertical shaded band indicates a plausible time span in which a putative magma ocean might have solidified, i.e., forming a stable solid surface; accumulation of craters on solid surfaces would occur only after that.

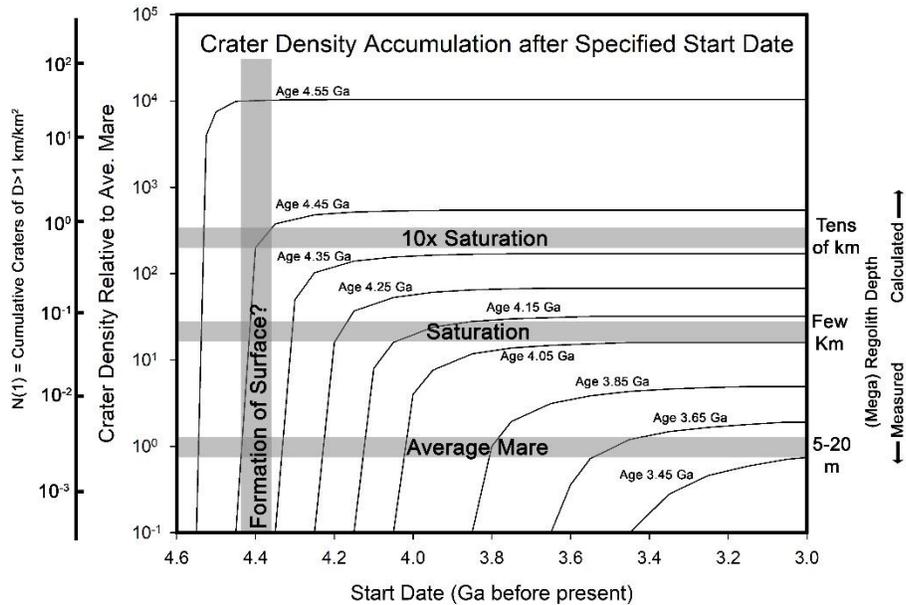

**Figure 3.** A plot of data from Table 2, showing accumulation of crater density as a function of time, during the putative early intense bombardment. Bottom of each curve (on abscissa) shows formation dates for surfaces, and the curves show subsequent crater accumulation on those surfaces. Data were initially plotted at 100 Ma intervals, leading to a sharp kink in each upper curve, ~100 Ma after the start date, due to short half-life of early impact rates; those kinks have been smoothed out by adding artificial data points to indicate a more realistic transition. (See text for discussion of the shaded bands and the three scales provided for the ordinate axis on left and right sides. See Table 2 footnote for definition of "average mare." The curve for 4.55 Ga is included to indicate the extreme impact rate if solid bodies existed at that time; the Moon probably did not yet exist.)

Figure 3 gives insights into a number of lunar properties and aspects of lunar history. For example:

- Crater SFD numbers formed on the earliest inner solar system surfaces, as represented by the 4.55 Ga and 4.45 Ga curves, are no longer measurable on surfaces today, because of extreme oversaturation. The earliest craters have been overlapped by later large craters, multi-ring basin impact structures, and basaltic mare lava flows.

- Surfaces formed prior to ~4.2 to ~4.3 Ga were intensely bombarded and super-saturated with craters, due to the very high early cratering rates, but are getting closer to saturation levels. Near saturation levels affect the survival of impact melts from upper kilometers that time.

- The youngest crater-saturated surfaces visible today should date from ~4.1 to ~4.2 Ga, consistent with Fig. 2a. The reduced saturation effects by this time allow for increasing survival of brecciated materials and impact melt inclusions from that time.

- Crater densities matching average mare surface appear around 3.5 to 3.6 as predicted from pre-Apollo crater counts and supported with dated Apollo samples.

# INTENSE EARLY BOMBARDMENT: CONSEQUENCES FOR MEGAREGOLITH EVOLUTION AND MEGAREGOLITH DEPTHS

Applying the information discussed so far, we investigate the evolution of megaregolith in the first 600 Myr of lunar history. Figure 4 introduces a technique for using the SFD curve of crater populations to calculate depth of pulverization, based on an earlier technique utilizing the isochron diagram (Hartmann 1980, 2005). The technique works for under-saturated, saturated, and over-saturated surfaces. For various SFD crater densities formed on such surfaces, we calculate the area covered by craters (per km$^2$) in each diameter bin, then add up the cumulative area starting with the largest bin and working toward smaller bins. As shown in Fig. 4, at each diameter D, this gives the total area covered by craters larger than D (and thus ejecting material deeper than some effective depth d excavated by such craters). At some sizes we reach a point size where the total area of larger craters is 100% of the area available. At 100% coverage, however, not all areas have been impacted and excavated, because of crater overlap. Thus, the 200% parameter is included to allow more realistically for such overlap. We will define our quantitative definition of d in a moment. We assume, as a first-order approximation, that when 100% to 200% of the surface is covered by craters larger than critical diameters designated as "$D_{100\%}$" and "$D_{200\%}$," respectively, then a characteristic depth of megaregolith is at least of the order ~$d_{100\%}$ to ~$d_{200\%}$. The effective depth at any diameter D must be less than the deepest part of the transient cavity, but deeper than the observed floor depth in the existing crater. Here we assume also that nearly all of the "effectively" excavated material is distributed elsewhere on the Moon. The depths affected by craters exceeding $D_{100\%}$ and $D_{200\%}$ are somewhat in question, since the transient cavity depths near the center of a crater are considerably greater than the final crater depths. What, then, is the depth actually affected by saturation coverage among craters of these sizes? Taking into account the effects mentioned so far, we assume conservatively that craters of the relevant diameters (some kilometers) churned and pulverized materials to effective depths equaling 1/3 of their diameter.

The depth of megaregolith, while averaging around the order of magnitude of ~$d_{100\%}$ to ~$d_{200\%}$, will vary from region to region, because stochastically-produced largest basin-forming impacts will blow away early surfaces and create new, localized, surfaces — which then begin to accumulate mare lava flows, new impacts, and new regolith. For example, the large, relatively late Imbrium basin must have blown away most earlier megaregolith (not to mention overlapping, obliterating, and ejecting material from earlier large craters and possible earlier, small multi-ring basins with their impact melts). At the same time, Imbrium's ejecta delivered such earlier megaregolith materials onto other regions (McGetchin et al. 1973; Haskin 1998; Hartmann 2003; Petro and Pieters 2008; Norman and Nemchin 2012; Norman et al. 2015). The giant basins thus create relatively thin spots in the megaregolith cover (possibly zero kilometers immediately after impact in some cases), while depositing the previous megaregolith material in a surrounding ejecta blanket. Another example of these effects may be present at South Pole-Aitken (SP-A) basin. Using GRAIL results, Besserer et al. (2014) state that "The SP-A region appears to possess a thinner low-density (porous) layer than the rest of the farside." This result suggests that that SP-A blew away a fair amount of megaregolith which had formed before the SP-A impact. Even cursory inspection of lunar photo images reveals that SP-A has a higher crater density and higher albedo than the major front-side maria. It has evidence of mare basaltic lava deposits, but those mare surfaces themselves have been heavily cratered, with the lava apparently pulverized and brightened by ray deposits. These statements suggest that SP-A and its lava flooding date from before most of better-preserved mare lava production dating from ~3.8 Ga to ~3.2 Ga, but nonetheless that a substantial thickness of megaregolith existed before the impact, to be blown away. This is consistent with our model, but inconsistent with the LHB paradigm, with its sparse cratering before 4.0 Ga.

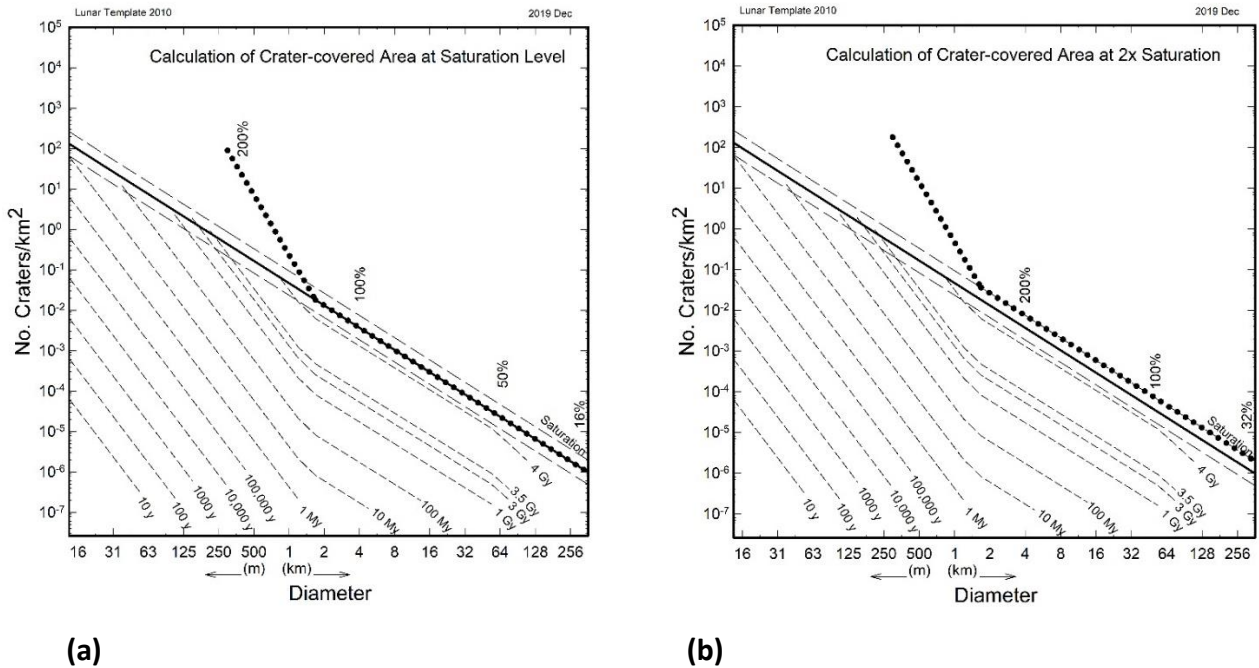

(a)  (b)

**Figure 4.** Estimated areas covered by craters larger than given diameters (indicated by vertical "%" figures. Fine-dashed SFD curves are isochrons showing observed crater SFDs as in Fig. 2a. Solid saturation line shows maximum observable crater densities reached on airless planetary bodies, based on empirical measurements. Note that when a given D bin reaches the saturation line it does not mean that craters larger than that D cover 100% of the Moon (see text). Heavy dotted curve in each figure shows actual number of impacts and craters formed, eventually rising above the saturation for small enough craters and large enough ages. Along the heavy dotted curves, calculations are made of cumulative area covered by craters larger than D, starting at lower right corner. (a): Isochron diagram for area just reaching saturation. Here, 100% of area is covered by craters larger than ~4 km, and 200% by craters larger than ~350 m. (b) The same but for a surface that is supersaturated by a factor 2. In this case, 100% of area is covered by craters larger than ~60 km and 200% covered by craters larger than ~5 km. The diagram illustrates how the sizes of craters excavating 100% and 200% of total area increases dramatically as super-saturation is reached.

Table 3 and Fig. 4 illustrate the technique described here.  We apply the technique starting with the saturation equilibrium curve, because it reflects the highest *observable* crater densities and thus suggests a minimal megaregolith depth in lunar highland saturated regions.  In Table 3, column 1 shows crater diameter and column 2 shows the calculated fractional area per km$^2$ covered by observed craters in each bin, as follows:

Fractional area covered in each bin = $\pi(D/2)^2$ N = 0.785ND$^2$     (1)

where
 D = crater diameter near mid-point in each bin
 N = no. craters/km$^2$ observed in each bin

This fractional area is then expressed as a percentage of the area in column 2 of Table 3.

The formula does not take into account that some craters overlap earlier craters.  This is a valid approach if we are interested in the total volume being excavated and ejected by *observable* craters.  But if we are interested in the total *surface* area that has been excavated, the 200% figure is better, since most of the area has been impacted at least once by craters larger than D$_{200\%}$.

Column 3 in Table 3 shows the cumulative percentage of area covered by all craters larger than D, starting with the largest craters at the bottom of the table.  At D >128 km there are poorer statistics because of the smaller number of large craters and basins, and the isochron curves are not plotted at still larger sizes for this reason.  Among the giant multi-ring basins at the largest-diameter end of the size-frequency distribution (SFD), at D >~360 km, there appears to be a continued turn-down at largest sizes, and we have added modest provisional percentage numbers in column 3 at D >360 km.  The areas covered, however, are not large enough to affect our conclusions significantly.  Columns 4, 5, and 6 add the same results for 2× saturation and 2.5× saturation

In Table 3, the important transitions at 100% and 200% coverage by impact craters are shown in bold face in columns 3, 5, and 6.  In this calculation, at D <1.4 km, we follow the well-known steep branch of the SFD upward into the region above the saturation line, not stopping at the saturation curve per se.  This is in order to include the effects of all impacts called for by the SFD at small sizes, whether we can see the resulting craters or not.  This steep branch (and presumably the larger craters of the large-D branch at D >1.4 km), include an uncertain fraction of secondary craters, which add to the churning and pulverization involved in megaregolith evolution.  [Our system of crater counts and crater chronometry includes *all* visible craters (excluding obvious clusters with rays), without trying to distinguish primaries from the remaining "field secondaries," since attempts to distinguish them visually are likely to be in error.  This system is practical, since nature can't produce large craters without producing secondaries, and can't produce secondaries without primaries, and on old surfaces with high crater densities, the addition of unrecognized scattered field secondaries in a single shower from a fresh distant primary will not have a drastic effect on the total crater density.]

Note that, as seen in Fig. 4a and mentioned above, the small-D end of the steep branch of the SFD is first to hit the saturation line. What appears to be sparse cratering as in lunar maria at large D thus produces supersaturation at small D creating a shallow regolith of roughly 5 to 15 meters depth, as detected in Apollo landings on lunar maria. But once saturation is approached at $D \gtrsim 1.4$ km, as indicated in Fig. 4a, the saturation line is reached virtually simultaneously at all larger crater diameters, causing an explosive growth in megaregolith.

Here we pause to ask whether these last statements apply to the earliest crater populations in the first few hundred million years. The problem is that if the early intense bombardment/accretion tail scenario is correct, we must ask if the SFD of crater populations at that time was the same as that observed in post- mare time, when the observed crater density is low, so that the SFD can be accurately measured. On earlier surfaces, dating from $\gtrsim 4.2$-4.3 Ga, the SFD is uncertain because supersaturation occurred, and details of the SFD shape are masked by supersaturation. However, Delbo et al. (2017) found that the collisional evolution of the asteroidal population (which produces the observed size distribution of asteroids) was already well developed prior to the dynamical instability among the giant planets, and possibly by the time of the formation of the Moon itself — and they concluded that its SFD at that time was already consistent with that of current asteroids (which, in turn, is at least roughly consistent with current craters). We thus see no firm observational or theoretical evidence that the slope of the SFD among the larger impactors in impactor populations before ~4.1-4.2 Ga departed from what we see recorded on unsaturated lunar surfaces today. Changes in typical velocity of crater-forming impactors are not expected to have an overwhelming effect on the shape of the SFD. The main point here is that the near-simultaneous approach to saturation by all craters of $D \gtrsim 1.4$ km caused a rapid transition from shallow regolith to deep megaregolith.

**Table 3.** Cumulative area covered by craters larger than diameter D among crater populations reaching the saturation equilibrium line, and also reaching 2.0× and 2.5× saturation equilibrium levels. Bold face type in columns 3, 5, and 6 emphasize conditions where ~100% and ~200% of area are covered by craters.

| Crater diameter D (km) marking boundaries of diameter bins | Percentage area covered by craters within specified bin when crater saturation is reached at D>1.4 km | Cumulative % of area in bin covered by all craters of diameter >D when saturation is reached (according to discussion in text, the youngest examples of saturated surfaces formed at 4.1 to 4.2 Ga) | Percentage area of craters/km² in bin when 2× crater saturation is reached at D>1.4 km | Cumulative % of area covered by craters of diameter >D at 2× saturation (according to Fig. 3, the youngest examples occur on surfaces formed at ~4.3 Ga) | Cumulative % of area covered by craters of diameter >D at 2.5× saturation (according to Fig. 3, the youngest examples would be >4.3 Ga) |
|---|---|---|---|---|---|
| 0.25 | | | | | |
| | 48% | **227%** | 96% | 456% | 567% |
| 0.353 | | | | | |
| | 30% | **179%** | 60% | 360% | 447% |
| 0.5 | | | | | |
| | 17% | 149% | 34% | 300% | 372% |
| 0.707 | | | | | |
| | 13% | 132% | 26% | 266% | 330% |
| 1 | | | | | |
| | 6% | 119% | 12% | 240% | 297% |
| 1.41 | | | | | |

| Crater diameter D (km) marking boundaries of diameter bins | Percentage area covered by craters within specified bin when crater saturation is reached at D>1.4 km | Cumulative % of area in bin covered by all craters of diameter >D when saturation is reached (according to discussion in text, the youngest examples of saturated surfaces formed at 4.1 to 4.2 Ga) | Percentage area of craters/km² in bin when 2× crater saturation is reached at D>1.4 km | Cumulative % of area covered by craters of diameter >D at 2× saturation (according to Fig. 3, the youngest examples occur on surfaces formed at ~4.3 Ga) | Cumulative % of area covered by craters of diameter >D at 2.5× saturation (according to Fig. 3, the youngest examples would be >4.3 Ga) |
|---|---|---|---|---|---|
| | 4% | 113% | 8% | 228% | 282% |
| 2 | | | | | |
| | 5% | 109% | 10% | 220% | 272% |
| 2.8 | | | | | |
| | 5% | **104%** | 10% | 210% | 260% |
| 4 | | | | | |
| | 4% | **99%** | 8% | **200%** | 247% |
| 5.66 | | | | | |
| | 6% | 95% | 12% | 192% | 237% |
| 8 | | | | | |
| | 6% | 89% | 12% | 180% | 222% |
| 11.3 | | | | | |
| | 6% | 83% | 12% | 168% | **207%** |

| Crater diameter D (km) marking boundaries of diameter bins | Percentage area covered by craters within specified bin when crater saturation is reached at D>1.4 km | Cumulative % of area in bin covered by all craters of diameter >D when saturation is reached (according to discussion in text, the youngest examples of saturated surfaces formed at 4.1 to 4.2 Ga) | Percentage area of craters/km² in bin when 2× crater saturation is reached at D>1.4 km | Cumulative % of area covered by craters of diameter >D at 2× saturation (according to Fig. 3, the youngest examples occur on surfaces formed at ~4.3 Ga) | Cumulative % of area covered by craters of diameter >D at 2.5× saturation (according to Fig. 3, the youngest examples would be >4.3 Ga) |
|---|---|---|---|---|---|
| 16 | | | | | |
| | 7% | 77% | 14% | 156% | **192%** |
| 22.6 | | | | | |
| | 7% | 70% | 14% | 142% | 175% |
| 32 | | | | | |
| | 6% | 63% | 12% | 128% | 157% |
| 45.3 | | | | | |
| | 9% | 57% | 19% | **116%** | 142% |
| 64 | | | | | |
| | 9% | 48% | 19% | **97%** | **120%** |
| 90.5 | | | | | |
| | 8% | 39% | 16% | 78% | **97%** |
| 128 | | | | | |

| Crater diameter D (km) marking boundaries of diameter bins | Percentage area covered by craters within specified bin when crater saturation is reached at D>1.4 km | Cumulative % of area in bin covered by all craters of diameter >D when saturation is reached (according to discussion in text, the youngest examples of saturated surfaces formed at 4.1 to 4.2 Ga) | Percentage area of craters/km$^2$ in bin when 2× crater saturation is reached at D>1.4 km | Cumulative % of area covered by craters of diameter >D at 2× saturation (according to Fig. 3, the youngest examples occur on surfaces formed at ~4.3 Ga) | Cumulative % of area covered by craters of diameter >D at 2.5× saturation (according to Fig. 3, the youngest examples would be >4.3 Ga) |
|---|---|---|---|---|---|
| 181 | 7% | 31% | 14% | 62% | 77% |
| 256 | 8% | 24% | 16% | 48% | 60% |
| 362 | 8% | 16% | 16% | 32% | 40% |
| 512 | ~5% | 8% | ~10% | ~16% | ~20% |
| 723 | ~2% | 3% | ~4% | ~6% | ~7.5% |
| 1024 | ~1% | 1% | ~2% | ~2% | ~2.5% |

What are the consequences of saturation and super-saturation for megaregolith evolution? At saturation level, as shown in Fig. 4a and Table 3, column 3, we see that $D_{200\%}$ is about 350 m, and $D_{100\%}$ is about 4 km. This would suggest effective regolith depths (allowing for the moment two significant figures) in the range of d ~120 m to ~1.3 km — somewhat deeper in some areas, and shallower in other areas, since craters larger than 4 km have more stochastic effects. These estimated regolith depths correspond to surfaces observed today but formed ~4.1 to ~4.2 Ga.

Now consider still older surfaces that are supersaturated with craters. Although the supersaturation cratering SFDs are hidden from the view of modern observers, Fig. 4b and Table 3 show how the accretion tail model would produce at least twice the level of the saturation curve on surfaces that formed as early as ~4.2-4.3 Ga — a date from which we have impact melt samples. In this case, as shown in column 4 of Table 3, craters of D >60 km would have covered 100% the surface, and craters of D >5 km would have covered 200% of the surface. These figures suggest that the megaregolith depth produced by 2× saturation would be (by our calculation) in the range of ~2 km to ~20 km. Note the explosive regolith growth within an interval of only ~100 Myr: While crater densities have increased by only factor 2 above saturation, the $d_{100\%}$ figure has increased from a depth of ~120 m to 2 km.

Because of the "explosive" growth effect, megaregolith depths increase rapidly as we consider slightly higher degrees of supersaturation on earlier (older) surfaces. On surfaces formed during a period earlier than ~4.3 Ga, when Table 3 indicates 2.5× saturation at $D_{100\%}$ = ~16 km to $D_{200\%}$ = ~90 km, megaregolith depth, as observed today, would be in the range 5 to 30 km (in areas not excavated by subsequent large impact basins – an effect mentioned above and discussed in the next section).

Interestingly, the megaregolith gravels appear frequently to have consolidated into coherent breccia material. Many of the Apollo highland breccias samples, perhaps excavated from depth, are strong, coherent rocks. Coherent materials may have arisen as a result of several different effects. First, impacts in the first hundred Myr or so may have penetrated primarily into molten materials, regionally stirring them and mixing them with shattered crustal materials, so that molten matrix materials may have helped solidify some breccias. Second, pressure effects may have helped cement the heated fragmented materials below some depth. Third, hypervelocity meteorite impacts at various scales produce glassy matrix material that helps to cement rocky fragments into coherent breccias. Wieczorek et al. (2013) and Spray (2016) have discussed such processes in more detail. To summarize, the depths of tens of kilometers, calculated here for impact-affected materials from the geometry of craters, should not necessarily be construed as occupied by loose material; the material may, however, be more porous and less dense than deeper plutonic layers unfractured by impacts. These results are consistent with GRAIL gravitational findings of low densities and higher porosities in the upper few kilometers of the lunar surface. For example, Wieczorek et al. (2013, p. 671) found densities of 2550 kg/m³ and average crustal porosity of 12% "to depths of at least a few kilometers." We suggest that these measurements refer to loosely consolidated and/or heavily fractured megaregolith, grading at depth into more coherent breccias and fractured crustal bedrock with very early plutonic igneous intrusions along some of the fractures.

To summarize this section so far, our discussions depart considerably from the "classic" terminal cataclysm/LHB models. To start with, a whole series of impacts at all scales, including a few $10^3$-km-scale basins, would have been forming as the putative magma ocean solidified. The end of magma ocean crystallization is believed to have happened by about 4.45 to 4.35 Ga (Elkins-Tanton et al. 2011; Zhu et al. 2019). Borg et al. (2011, 2015) describe an age of 4.360±0.003 Ga for a ferroan anorthosite sample 60025 and discuss it as marking a possible magma ocean crystallization age. However, they also refer to three ferroan anorthosites dated by other authors at 4.43±0.03, 4.47±0.07, and 4.53±0.12 Ga, which could record earlier impact and recrystallization events. Numbers of impact melt clasts, dated at ~4.35 Ga, offer vestiges of one or more large impact structures during that period (Nemchin et al. 2008, Grange et al. 2013, and White et al. 2020; see also Hartmann 2019 for review). Figures 1, 2a, and 3 of our present paper imply very intense bombardment, including multiple impacts of basin-forming magnitude, along with extreme crater oversaturation, during the suggested periods of magma ocean presence. In contrast, the LHB paradigm, from its original version (Wetherill 1975, 1977; Ryder 1990) with its period of several hundred million years of negligible impacts before 4.0 Gyr, to the re-defined versions (e.g., Bottke 2012, with LHB reduced in intensity to a surge from 4.1 to 2.5 Ga) all involved a period of relatively low bombardment during the suggested periods of magma ocean presence.

Our accretion tail scenario, with early intense bombardment during the putative magma ocean's lifetime and solidification, would produce radically different petrologic results than the LHB models. Some early discussions of the solidification of the putative lunar magma ocean, under conditions pictured in LHB models, assumed the process starting with smooth-layered stratigraphy evolving in the absence of large impacts, with density-controlled layering according to the Bowen reaction series. For example, Walker (2009, p. 109) stated that "…the lunar mantle is likely stratified resulting from magma ocean crystallization and little subsequent mixing." The accretion tail model of impact history, together with early work on planetary accretion (e.g., Safronov 1972) and our analysis here, indicate that during accretion of the Moon and the final solidification of the magma ocean, the uppermost mantle of the Moon and/or magma ocean could not have solidified in a smoothly "stratified" way with "little subsequent mixing." Addressing the date of solidification for the magma ocean, Borg et al. (2011) considered their zircon-derived 4.36 Ga date to be surprisingly young and concluded "much of the lunar crust may have been produced by non-magma-ocean processes, such as serial magmatism." In contrast to LHB scenarios, our proposed early intense bombardment allows an explanation for complex crustal evolution. serial magmatism, and HSE anomalies. Large-scale basin impacts, would have broken up newly forming surface crustal layers in many regions and ejected masses of deeper material into widespread megaregolith layers, as well as triggering magmatic eruptions. Impact-fracturing and impact-stirring of deep magmas may have been a factor in allowing HSE to collect in the lunar interior. Magma at depth from would have been stirred and redistributed during the period from, say, 4.35 Ga to ~4.1 Ga.

Figure 2b, representing the intense impact regime about 4.35 Ga, suggests impact basins of >300 km diameter forming stochastically at intervals on the order of ten million years in random lunar locations at that time. Each one would have excavated to depths of many tens of km and dumped ejected fragmental material over wide areas of the Moon. Consistent with this, Wieczorek and Phillips (1999) suggested that large impacts may have disturbed the crustal structure being formed by magma ocean crystallization. Gross and Joy (2016) and Pernet-Fisher and Joy (2016) also discussed effects of impacts on magma ocean evolution and crustal formation. Rolf et al. (2017) suggested that accumulating ejecta deposits (megaregolith) from impact basins scattered over time may have an insulating effect, influencing the rate of heat flow from the interior.

## CONSEQUENCES OF LARGEST EARLY IMPACTS
### *VIS À VIS* HETEROGENEITY OF THE MEGAREGOLITH

The formation of multi-ring basins approaching 1000 km diameters, scattered in space and time during the first 600 Myr, created localized effects on the depth of megaregolith (which included mixed layers of local and distant basin ejecta). Such giant basins also affect the observable survival of earlier, nearby basins, by obliterating parts of rim structures and/or burying parts of them in ejecta. The accretion tail model with early intense bombardment thus casts the South Polar Aitken Basin and the controversial Oceanus Procellarum in a new light. Figures 1, 2a, and Table 1 suggest that under our proposed cratering rates before ~4.3 Ga, several giant impact basins, in the $10^3$ km diameter (outer ring) range would probably have formed.

The Procellarum area was once considered to be a candidate for such a primordial impact basin, and this hypothesis gains plausibility in the context of our discussion. The GRAIL team discovered a 3000-km diameter "ring" of linear gravitational anomalies under the outer part of Oceanus Procellarum, interpreted as intrusion-filled fractures — which might seem to support that Procellarum is a form of a multi-ring impact feature. They suggested, however, that since these features comprise about five or six linear segments, they are not related to the "rings" of faults around multi-ring impact basins (Andrews-Hanna et al. 2014). Thus they argued against an impact origin for Procellarum. Figure 5, however, shows ~30-40-degree linear segments in some rings around the Orientale impact basin. Figure 6 shows similar linear segments in the ring structures around the Humboldtianum impact basin. Note that if a truly circular mass of material along a crater rim is trying to slide downward into a basin interior along a conical fault, it is inhibited from sliding because it is buttressed by its own material. Linear normal fault segments make it easier for such material to slip, in segments, down into the crater interior. Furthermore, a Procellarum-scale impact involves restructuring the spherical curvature of the Moon, and we have little knowledge of how a spherical planetary body's brittle and/or plastic lithospheric layers adjust to world-scale impact craters. Our conclusion is that linear subsurface features at Procellarum do not argue against an impact origin for that large lunar feature. Zhu et al. (2017) also argue in favor of a collisional origin of Procellarum because this event would help explain the dichotomy in crustal thickness between the near-side and far-side of the Moon.

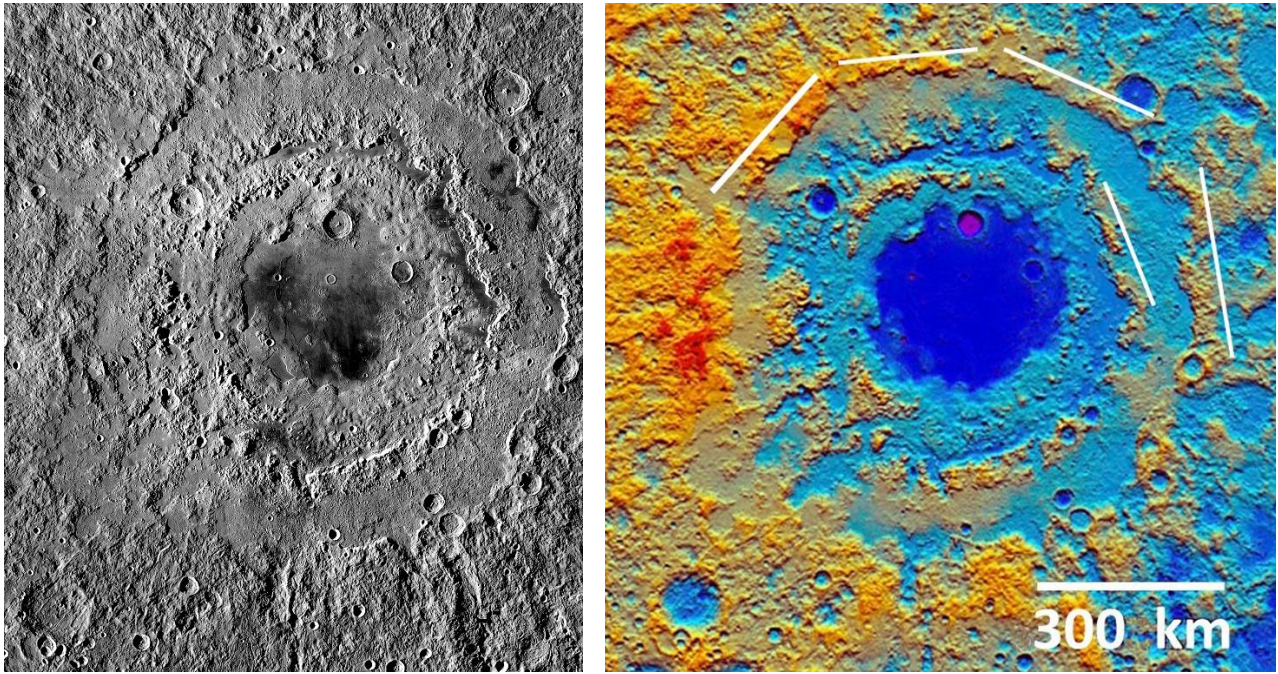

(a)                                         (b)

**Figure 5.** Orientale basin, in an orbital photo (a) and a topographic map (b) showing linear segments in the surrounding rings.  Three linear segments are pronounced along the north (top) third of the outer (930 km diameter) ring, especially at top of the topographic map.  The outer ring and next inner ring are formed by linear segments on ENE side.   Note in 5(a) both of those segments have a narrow strip of dark lava at the inner foot of the scarp, suggesting a fault plane that allowed magma to intrude along the fault and then extrude onto the surface.  An outer ring of diameter 1300 km has also been mapped, but is not easily visible here.  (a)  NASA Lunar Reconnaissance Orbiter Camera WAC image [NASA/Goddard Space Flight Center].  (b)  Laser altimeter "LOLA" topographic map from NASA/Goddard Space Flight Center.)

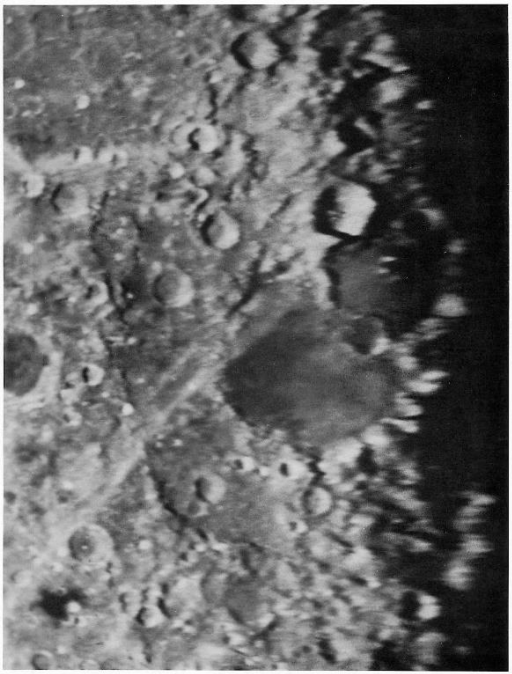 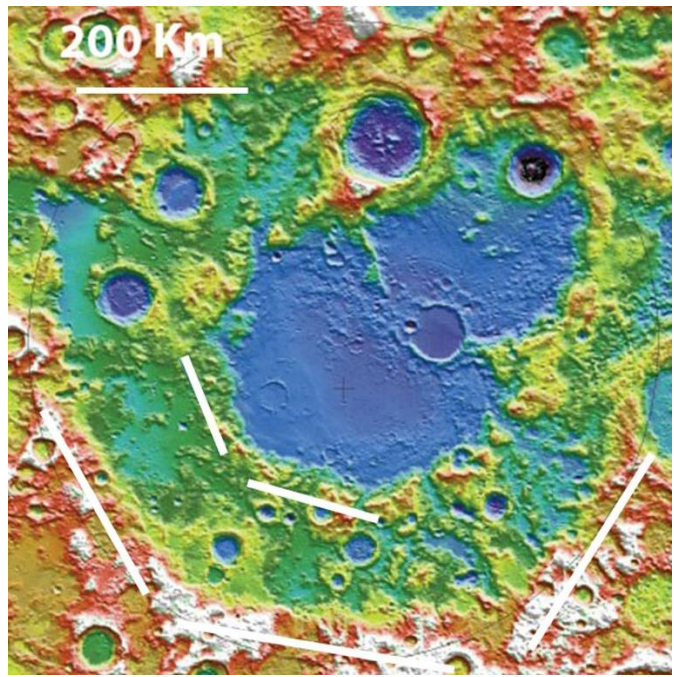

(a)                  (b)

**Figure 6.** Humboldtianum multi-ring basin. Photo (a) and topographic map (b) reveal that the southern half of the outer ring (boundary between orange highlands surrounding and green, depressed interior), shows three clear linear segments. Diameter of outer ring is ~620 km. NE half portion of that ring and the basin are overlain by a large crater. (a) Discovery rectified photo by WKH, from Hartmann and Kuiper (1962). This early image is shown in order to illustrate that these linear features have been apparent since the early 1960s. (b) NASA; "LOLA" topographic map: NASA/Goddard Space Flight Center.

       Another consequence of the accretion tail model with early intense bombardment is that the earliest, extremely high impact rates such as 10,000 times the present rate, go beyond the rates typically contemplated in the terminal cataclysm/LHB model. We have already mentioned that the earliest basins would have been beaten into presently undetectable states. Figure 7 illustrates the effectiveness of that process. It shows a vivid example of an older basin, Serenitatis, whose rim has been broken into linear ridges and partly demolished by ejecta or faulting associated with the younger, neighboring Imbrium basin. The mountains of the Serenitatis rim are laced by ridges and valleys structured not radial to Serenitatis but radial to Imbrium, whose center is ~980 km away, beyond the lower edge of the frame. The destruction here is also ~270 to ~390 km beyond the Apennine rim of the Imbrium basin. The precise mechanism for this reshaping of the Serenitatis rim by the Imbrium impact is still not certain. Tectonic fractures caused by the Imbrium impact, with collapses during lava flooding of the area, seem a possibility. A second possibility might be gouging by local grazing, low-angle rock masses in ejecta ejected from Imbrium, possibly embedded in a somewhat volatilized mass like a base surge. The second mechanism seems perhaps less preposterous when one considers that the "mountain ranges" creating basin rims are, to some extent, piled-up, ejected megaregolith gravels. Whatever the cause, the degree of Serenitatis destruction by Imbrium is remarkable — and all the more so since (as discussed in more detail by Hartmann 2019, pp. 19-20, Table 1) there is at least one case where researchers relying on radiometric dating of Apollo samples listed Serenitatis as about one error bar *younger than Imbrium*. (Most proposed dating of Serenitatis

put it within an error bar to two *older* than Imbrium.) Such young dates reported for Serenitatis may, however, have involved the Imbrium ejecta that modified the Serenitatis rim (see Hartmann 2019 for more detailed discussion).

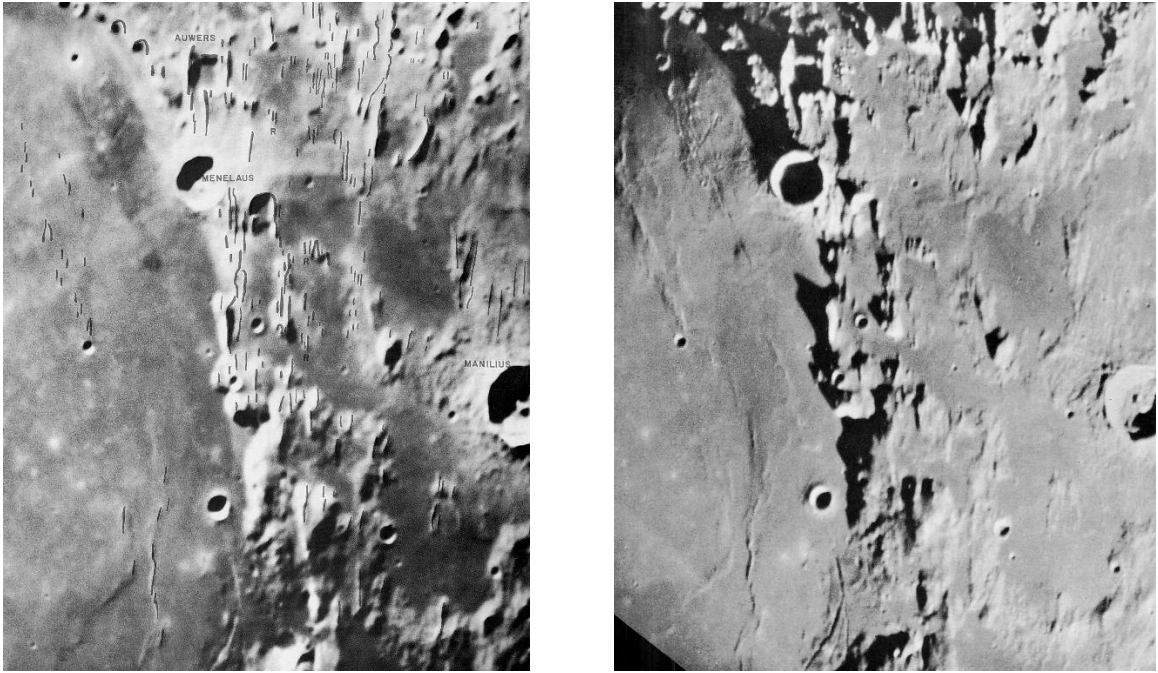

(a)                                                                 (b)

**Figure 7.** Two views of south rim of Serenitatis basin (rugged terrain from bottom to top along center of image), showing degradation by linear structural features radial to Imbrium basin (centered ~270 km off the bottom of the image; north at left). The pattern of ridges radial to Imbrium extends as far as the Apollo 16 site, where Imbrium-age material was collected. Bottom to top length of each image ~350 km. (Photos by WKH, during Lunar Rectified Atlas project ca. 1962, University of Arizona, Lunar and Planetary Lab.)

## CONSEQUENCES OF MEGAREGOLITH EVOLUTION *VIS-À-VIS* ROCK SAMPLES COLLECTED ON PRESENT-DAY LUNAR AND PLANETARY SURFACES

The evolution of the megaregolith, and its current properties, have important effects on the rocks and soil samples collected on the present-day surface of the Moon, not to mention other planetary surfaces. To take impact melt samples first, the high formation rate of early basins and large craters had important effects on any impact melts formed in the first few hundred Myr. The large, multi-ring basins are believed to contribute most of the volume of lunar impact melts (Artemieva and Shuvalov 2008), but the exact disposition and depth of impact melts within basins appear to be uncertain. For example, based on terrestrial observations and theoretical modeling, Cintala and Grieve (1998) depict impact melt sheets as layers much thinner than the transient cavity basin depth, spread across the basin interior.  Artemieva and Shuvalov (2008) and others suggest more concentration in deeper plugs under the middle of the crater.  Impact melt plugs as deep as 20 km have been suggested under Imbrium.  It is implicit in either case that impact melts are created and distributed in more or less *surficial* layers having to do with impacts and ejecta blankets. We have shown, however, that in the accretion tail model, creation of the initial megaregolith in the first few hundred Myr involved pulverization to depths of tens of kilometers, probably deeper than most impact melts.  As also shown in Fig. 3, surfaces and near-surface structures formed around 4.15 Ga are only slightly supersaturated.  By about 4.0 or 3.9, near-surface structures such as impact melt lenses in the upper few kilometers are more likely to survive.  Furthermore, as indicated in Fig. 3 and Fig. 2a, structures formed around 4.0 Ga no longer become saturated with large craters.  In short, the accretion tail scenario, contrary to terminal cataclysm/LHB models, explains why impact melts older than ~4.1 Ga are hard to find, while increasing numbers of impact melts are found from  ~4.0 Ga to 3.9 Ga.  The number of impact melts decreases after that because the impact rate itself was declining. In this view, the unique aspect of 3.9 Ga is not a sudden narrow spike in impact rate, but rather that this date marks the beginning of the period when massive formations on the scale of impact melt lenses will survive — not to mention that the Imbrium and Orientale basins formed around that time.

But what about the fact that primordial crustal samples date back as far as 4.44-4.45 Ga?  Why weren't they also destroyed?  The answer lies in the fact that the base of the megaregolith, by definition, lies as deep as the deepest pulverization and fragmentation, reaching into more intact primordial rock layers.  According to our scenario of early intense bombardment and accretion tail, the megaregolith reached tens of kilometers depth.  Richardson and Abramov (2020) emphasize different layers in the megaregolith, with a similar conclusion.  They picture a surface dusty regolith layer typically about 20 m deep, underlain by an "upper megaregolith" from ~20 m to a depth of 1 to 3 km, consisting of depositional layers of "brecciated and/or melted material…characterized by the transport and deposition of material via either transient crater gravitational collapse or impact ejecta." Under the "upper regolith," they picture a "lower megaregolith" layer, running from ~1-3 km at the top to "about 20-25 km," consisting of bedrock that has been fractured in place.  In all such models there is a semi-infinite supply of primordial, lower crustal or upper mantle plutonic rocks below the bottom of the megaregolith.  Thus, the largest modern craters, whose transient cavities exceed roughly 20 km depth, excavate samples of primordial crustal or upper mantle rocks and scatter them on the surface.

This excavation process is aided by additional aspect of megaregolith evolution:  It is heterogeneous in the sense that the biggest basin impacts blow away much of the accumulated megaregolith under their floors, creating "thin spots."  Those areas were then coated with lavas, but even modest-sized craters on floors of basins can penetrate the lavas and eject ancient, deep-seated rocks.  Craters with bright ray systems, associated with Imbrium and postdating mare lava plains, such as Autolycus (diameter D = 81 km) and Aristillus (D = 55 km) (both inside the Imbrium's Apennine ring) are good examples.  Copernicus (D = 93 km) and Eratosthenes (D = 59 km), somewhat outside that ring, are also candidates for ejecting early crustal samples.

Once a rock has been expelled onto the lunar surface, how long does it last? This question was not much considered when the Apollo missions were planned and the first samples were collected, because the Moon was thought to lack erosive processes. However, a number of diverse studies show that lunar rock samples typically have cosmic ray exposure ages of tens of Ma to a few hundred Ma, and meter-scale rocks typically erode due to the "sandblasting" flux of small meteorites with a few hundred Ma. Moreover, craters of D <12 km (depth ≲2 km) typically produce rocky ejecta blankets in lunar maria (from lava layers) but not so much in lunar highlands (where the subsurface is mostly pulverized). For discussion of such effects, see Basilevsky et al. 2013; Ghent et al. 2014; Costello et al. 2018; Hartmann 2019. Thus, in collecting lunar rock samples, we are entirely at the mercy of excavation processes in the megaregolith; megaregolith properties "filter" what we can collect on the surface.

**FUTURE WORK**

We regard our current work as a demonstration of concept. In the past there has been some separation of sub-disciplines in planetary science. Research in sample petrology, radiometric dating, impact crater studies, numerical modeling of interplanetary impact fluxes have typically involved different research groups, different meetings, different sessions at large annual meetings, and different journals. One area of fruitful future research may attempt to combine better the insights from (1) dynamical models of accretion, planetesimal collisions, and planetesimal fragmentation events, (2) the resulting geological consequences for lunar and planetary cratering rates as a function of time, (3) studies of the processes of excavation and deposition of impact debris. We note that such work will apply to more or less airless bodies throughout at least the asteroid belt and inner solar system, including Mars. At the same time, we caution that to do more detailed modeling of megaregolith evolution we need better quantitative information on still-vague parameters such as (1) extent and depth of fracturing and pulverization of rock during impacts of various sizes; (2) velocity distributions for the different kinds of meteorites hitting the Moon at any given time; (3) cross-section and volume of transient cavities during impacts of various sizes, especially multi-ring basins; (4) cross-section and depth of impact melt lenses on or under the floors of multi-ring impact basins; (5) deposition depths, compositions, and homogeneity of structure in ejecta blankets as a function or distance from crater center for craters of various sizes (especially looking at sources of ejected KREEP materials and uniqueness of KREEP material to the Imbrium region); (6) degree and causes of cementation of fractured and pulverized megaregolith materials as a function of depth; (7) nature of rocks excavated from the base of the megaregolith by the largest recent craters in terms of our access to the original lunar crust or the upper mantle; (8) survival times of excavated rocks as a function of sizes on the lunar (or planetary) surfaces (i.e., erosion histories of samples accessible on the surface.

## CONCLUSIONS

The concepts of "early intense bombardment" and the "accretion tail" dynamical model (Morbidelli et al. 2018) provide a coherent view of age distributions of radiometric dates of lunar primordial crustal rocks vs. impact melts, GRAIL data on crustal structures, and early evolution of lunar features and multi-ring basin morphologies.  They also clarify phenomena not adequately explained by the terminal cataclysm/LHB paradigm.  Among the implications of our study:

- The solidification of the magma ocean was affected by extremely high cratering rates during that early era.  This prevented "smooth," layered deposits based on Bowen reaction series in at least some areas, and may help explain complex petrology among the most ancient samples from various regions of the present-day Moon.
- If a surface formed (or attempted to form) by solidification of a magma ocean during the extreme bombardment as early as >4.4 to 4.35, it would have experienced supersaturation of multi-kilometer diameter crater size-frequency distributions by about 4.35 to 4.3 Ga.
- Surviving remnants of such early surfaces, i.e., between the largest impact basins, would have continued to accumulate even more craters, so that early crater densities would have gone into supersaturation, probably by factors at least 2.5 to 10, resulting in pulverization of the earliest impact melts.  This result differs from predictions of the terminal cataclysm/LHB models, and helps explain the dearth of impact melt clasts from before ~4.35 Ga.
- Continuing bombardment in the first few hundred Myr of crust existence produced reworking and redistribution of materials in upper megaregolith layers, consistent with GRAIL measurements of low density and high porosity in the upper kilometers, Richardson and Abramov's (2020) modeling, and age distribution of earliest known impact melts.
- The surface today would display localized regions destroyed and resurfaced by multi-ring basin-scale impacts, typically filled with basaltic lavas, all in various stages of degradation by subsequent cratering.  This scenario is consistent with the presence of the large, very degraded South Polar Aitken basin, and with Oceanus Procellarum being a very early giant impact feature.
- Subsurface linear features comprising a "ring" around Oceanus Procellarum are not a compelling argument against Procellarum being an impact feature.
- The oldest lava-plain formations, for example, in South Polar Aiken, are covered by a veneer of high-albedo ejecta and overlapping rays, and were often mapped in the pre-Apollo and early post-Apollo eras as "highlands."   They may, however, have quite different megaregolith composition (and other properties) than highland regions that survive from earliest crustal formation
- The supersaturation-pulverization of the upper kilometers would have comminuted the impact melt lenses produced in the surface layers of the early impact basins, explaining the presence of impact melts from ~ 4.1 to 4.35 Ga as fine-scale clasts in lunar breccias.
- Surviving surfaces formed around 4.1-4.2 Ga have reached at least 1.0× saturation crater density.

- Surfaces formed after about 4.0-3.9 Ga are not quite saturated with craters, and experienced much lower regolith production rates than earlier surfaces. Survival times of rocks in surface layers rapidly increase at this time. This explains the survival of many impact melt rock samples after ~4.0-~3.9 Ga as contrasted with few before that time.
- The blasting away of megaregolith layers by a few large, late basin scale impacts (especially Imbrium) created thin spots in the megaregolith where recent Copernicus-scale and Autolycus-scale craters can eject primordial crustal rocks from the base of the local megaregolith, which in many instances is covered by mare basalt layers. This explains the presence of primordial plutonic crustal rocks in present-day surface sample collections (in spite of the paucity of impact melt rocks before ~4.0 Ga).
- Many concepts in this paper, regarding megaregolith evolution and sample collection apply not only to the Moon but to all relatively airless planetary bodies in the inner solar system (Mercury, moons, asteroids, and even Mars), and at least into the main asteroid belt, and probably beyond, depending on the histories of asteroid/comet impact rates in the outer solar system.
- We are still in an era where first-significant-figure issues remain to be considered, even while assessing certain processes with numerical models citing two- and three-significant-figures.

## ACKNOWLEDGMENTS

This paper was written mostly during visiting scientist invitations at the International Space Science Institute in Bern, Switzerland; thanks to ISSI for hospitality. Additional thanks to the Planetary Science Institute, Tucson, for emeritus office space, and Elaine Owens (at PSI) for invaluable help with editorial issues and Dan Berman for critiques and appreciated help in preparing the figures. We thank Katherine Joy, Stephanie Werner, Axel Wittmann, Mark Wieczorek, and Nicole Zellner for very helpful comments on drafts of this paper.